# Quantifying social group evolution

Gergely Palla[1], Albert-László Barabási[2] and Tamás Vicsek[1,3]

[1]Statistical and Biological Physics Research Group of HAS, Pázmány P. stny.1A, H-1117 Budapest, Hungary,
[2]Center for Complex Network Research and Department. of Physics, University of Notre Dame, IN 46566, USA.
[3]Department. of Biological Physics, Eötvös University, Pázmány P.stny.1A, H-1117 Budapest, Hungary.

**The rich set of interactions between individuals in the society [1,2,3,4,5,6,7] results in complex community structure, capturing highly connected circles of friends, families, or professional cliques in a social network [3,7,8,9,10]. Thanks to frequent changes in the activity and communication patterns of individuals, the associated social and communication network is subject to constant evolution [7,11,12,13,14,15,16]. Our knowledge of the mechanisms governing the underlying community dynamics is limited, but is essential for a deeper understanding of the development and self-optimisation of the society as a whole [17,18,19,20,21,22]. We have developed a new algorithm based on clique percolation [23,24], that allows, for the first time, to investigate the time dependence of overlapping communities on a large scale and as such, to uncover basic relationships characterising community evolution. Our focus is on networks capturing the collaboration between scientists and the calls between mobile phone users. We find that large groups persist longer if they are capable of dynamically altering their membership, suggesting that an ability to change the composition results in better adaptability. The behaviour of small groups displays the opposite tendency, the condition for stability being that their composition remains unchanged. We also show that the knowledge of the time commitment of the members to a given community can be used for estimating the community's lifetime. These findings offer a new view on the fundamental differences between the dynamics of small groups and large institutions.**

The data sets we consider contain the monthly roster of articles in the Los Alamos cond-mat archive spanning 142 months, with over 30000 authors [25], and the complete record of phone-calls between the customers of a mobile phone company spanning 52 weeks (accumulated over two week long periods), and containing the communication patterns of over 4 million users. Both type of collaboration events (a new article or a phone-call) document the presence of social interaction between the involved individuals (nodes), and can be represented as (time-dependent) links. The extraction of the changing link weights from the primary data is described in the Supplementary Information. In Fig.1a-b we show the local structure at a given time step in the two networks in the vicinity of a randomly chosen individual (marked by a red frame). The communities (social groups represented by more densely interconnected parts within a network of social links) are colour coded, so that black nodes/edges do not belong to any community, and those that simultaneously belong to two or more communities are shown in red.

The two networks have rather different local structure: due to its bipartite nature, the collaboration network is quite dense and the overlap between communities is very significant, whereas in the phone-call network the communities are less interconnected and are often separated by one or more inter-community nodes/edges. Indeed, while the phone record captures the communication between two people, the publication record assigns to all individuals that contribute to a paper a fully connected clique. As a result, the phone data is dominated by single links, while the co-authorship data has many dense, highly connected neighbourhoods. Furthermore, the links in the phone network correspond to instant communication events, capturing a relationship as it happens. In contrast, the co-authorship data records the results of a long term collaboration process. These fundamental differences suggest that any common features of the community evolution in the two networks represent potentially generic characteristics of community formation, rather than being rooted in the details of the network representation or data collection process.

The communities at each time step were extracted using the Clique Percolation Method [23,24] (CPM). The key features of the communities obtained by the CPM are that (i) their members can be reached through well connected subsets of nodes, and (ii) the communities may overlap (share nodes with each other). This latter property is essential, since most networks are characterised by overlapping and nested communities [6,23]. As a first step, it is important to check if the uncovered communities correspond to groups of individuals with a shared common activity pattern. For this purpose we compared the average weight of the links inside communities, $w_c$, to the average weight of the inter-community links, $w_{ic}$. For the co-authorship network $w_c/w_{ic}$ is about 2.9, while for the phone-call network the difference is even more significant, since $w_c/w_{ic} \approx 5.9$, indicating that the intensity of collaboration/communication within a group is significantly higher than with contacts belonging to a different group [26,27,28]. While for co-authors the quality of the clustering can be directly tested by studying their publication records in more detail, in the phone-call network personal information is not available. In this case the zip-code and the age of the users provide additional information for checking the homogeneity of the communities. According to Fig.1c the $<n_{real}>/<n_{rand}>$ ratio is significantly larger than 1 for both the zip-code and the age, indicating that communities have a tendency to contain people from the same generation and living in the same neighbourhood ($<n_{real}>$ is the size of the largest subset of people having the same zip code averaged over time steps and the set of available communities, while $<n_{rand}>$ represents the same average but with randomly selected users). It is of specific interest that $<n_{real}>/<n_{rand}>$ for the zip-code has a prominent peak at $s \approx 35$, suggesting that communities of this size are geographically the most homogeneous ones. However, as Fig.1d shows, the situation is more complex: on average, the smaller communities are more homogeneous in respect of both the zip-code and the age, but there is still a noticeable peak at $s \approx 30$-35 for the zip-code. In summary, the phone-call communities uncovered by the CPM tend to contain individuals living in the same neighbourhood, and having a comparable age, a homogeneity that supports the validity of the uncovered community structure. Further support is given in the Supplementary Information.

The basic events that may occur in the life of a community are shown in Fig.1e: a community can grow or contract; groups may merge or split; new communities are born while other ones may disappear. We have developed a method for the appropriate matching of the evolving communities from the information available for relatively distant points in time only (see Methods).

After determining the dynamically changing community structure, we first consider two basic quantities characterising a community: its size $s$ and its age $\tau$, representing the time passed since its birth. $s$ and $\tau$ are positively correlated: larger communities are on average older (Fig.2a. Next we used the auto-correlation function, $C(t)$, to quantify the relative overlap between two states of the same community $A(t)$ at $t$ time steps apart:

$$C_A(t) \equiv \frac{|A(t_0) \cap A(t_0 + t)|}{|A(t_0) \cup A(t_0 + t)|}, \quad (1)$$

where $|A(t_0) \cap A(t_0 + t)|$ is the number of common nodes (members) in $A(t_0)$ and $A(t_0+t)$, and $|A(t_0) \cup A(t_0 + t)|$ is the number of nodes in the union of $A(t_0)$ and $A(t_0+t)$. Fig.2b shows the average time dependent auto-correlation function for communities born with different sizes. The data indicate that the collaboration network is more "dynamic" *(<C(t)>* decays faster). We also find that in both networks, the auto-correlation function decays faster for the larger communities, showing that the membership of the larger communities is changing at a higher rate. On the contrary, small communities change at a smaller rate, their composition being more or less static. To quantify this aspect of community evolution, we define the *stationarity* $\zeta$ of a community as the average correlation between subsequent states:

$$\zeta \equiv \frac{\sum_{t=t_0}^{t_{max}-1} C(t, t+1)}{t_{max} - t_0 - 1}, \quad (2)$$

where $t_0$ denotes the birth of the community, and $t_{max}$ is the last step before the extinction of the community. Thus, $1 - \zeta$ represents the average ratio of members changed in one step.

We observe an interesting effect when investigating the relationship between the lifetime $\tau^*$ (the number of steps between the birth and disintegration of a community), the stationarity and the community size. The lifetime can be viewed as a simple measure of ``fitness'': communities having higher fitness have an extended life, while the ones with small fitness quickly disintegrate. In Fig.2c-d we show the average life-span $<\tau^*>$ (colour coded) as a function of the stationarity $\zeta$ and the community size $s$ (both $s$ and $\zeta$ were binned). In both networks, for small community sizes the highest average life-span is at a stationarity value very close to one, indicating that for small communities it is optimal to have static, time independent membership. On the other hand, the peak in $<\tau^*>$ is shifted towards low $\zeta$ values for large communities, suggesting that for these the optimal regime is to be dynamic, i.e., a continually changing membership.

To illustrate the difference in the optimal behaviour (a pattern of membership dynamics leading to extended lifetime) of small and large communities, in Fig.3. we show the time evolution of four communities from the co-authorship network. As Fig.3. indicates, a typical small and stationary community undergoes minor changes, but lives for a long time. This is well illustrated by the snapshots of the community structure, showing that the community's stability is conferred by a core of three individuals representing a collaborative group spanning over 52 months. In contrast, a small community with high turnover of its members has a lifetime of nine time steps only (Fig.3b). The opposite is seen for large communities: a large stationary community disintegrates after four time steps (Fig.3c). In contrast, a large non-stationary community whose members change dynamically, resulting in significant fluctuations in both size and the composition, has quite extended lifetime (Fig.3d).

The quite different stability rules followed by the small and large communities raise an important question: could the inspection of a community itself predict its future? To address this issue, for each member in a community we measured the total weight of this member's connections to outside of the community ($w_{out}$) as well as to members belonging to the same community ($w_{in}$). We then calculated the probability that the member will abandon the community as a function of the $w_{out}/(w_{in}+w_{out})$ ratio. As Fig.4a shows for both networks, if the relative commitment of a user to individuals outside a given community is higher, then it is more likely that he/she will leave the community. In parallel, the average time spent in the community by the nodes, $<\tau_n>$, is a decreasing function of the above ratio (Fig.4a inset). As Fig.4a shows, those with the least commitment have a quickly growing likelihood of leaving the community. Taking this idea from individuals to communities, we measured for each community the total weight of links (a measure of how much a member is committed) from the members to others, outside of the community ($W_{out}$), as well as the aggregated link weight inside the community ($W_{in}$). We find that the lifetime of a community decreases for large $W_{out}/(W_{in}+W_{out})$ ratios (Fig.4b inset). However, an interesting observation is that, while the lifetime of the phone-call communities for moderate levels is relatively insensitive to outside commitments, the lifetime of the collaboration communities possesses a maximum at intermediate levels of inter-collaborations (collaboration between colleagues who belong to different communities). These results suggest that a tracking of the individual's as well as the community's relative commitment to the other members of the community provides a clue for predicting the community's fate.

In summary, our results indicate the significant difference between smaller collaborative or friendship circles and institutions. At the heart of small cliques are a few strong relationships, and as long as these persist, the community around them is stable. It appears to be almost impossible to maintain this strategy for large communities. Our calculations show that the condition for stability of large communities is continuous changes allowing that after some time practically all members are exchanged. Such loose, rapidly changing communities are reminiscent of institutions, that can continue to exist even after all members have been replaced by new members. For example, in a few years most members of a school or a company could change, yet the school and the company will be detectable as a distinct community at any time during its existence.

**METHODS**

**Locating communities**. In the CPM method a community is defined as a union of all *k-cliques* (complete sub-graphs of size *k*) that can be reached from each other through a series of adjacent *k*-cliques (where adjacency means sharing *k-1* nodes) [24,29]. When applied to weighted networks, the CPM has two parameters: the *k*-clique size *k*, (in Fig.1a-b we show the communities for *k*=4), and the weight threshold *w\** (links weaker than *w\** are ignored). The criterion for selecting these parameters is discussed in the Supplementary Information.

**Identifying evolving communities**. The basic idea of the algorithm developed by us to identify community evolution is shown in Fig.1f. For each consecutive time steps *t* and *t+1* we construct a joint graph consisting of the union of links from the corresponding two networks, and extract the CPM community structure of this joint network (we thank I. Derényi for pointing out this possibility). Any community from either the *t* or the *t+1* snap-shot is contained in exactly one community in the joint graph, since by adding links to a network, the CPM communities can only grow, merge or remain unchanged. Thus, the communities in the joint graph provide a natural connection between the communities at *t* and at *t+1*. If a community in the joint graph contains a single community from *t* and a single community from *t+1*, then they are matched. If the joint group contains more than one community from either time steps, the communities are matched in descending order of their relative node overlap (see the Supplementary Information).

**Acknowledgements** We thank I. Derényi for useful suggestions and G. Szabó for his assistance with the mobile communication dataset. G.P and T.V are supported by grants from OTKA Nos: F047203 and T034995; A.L.B is supported by the James S. McDonnell Foundation and the National Science Foundation ITR DMR-0426737 and CNS-0540348 within the DDDAS program.

**Supplementary Information** is linked to the online version of the paper at www.nature.com/nature

## Figure legends

**Figure 1. Structure and schematic dynamics of the two networks considered**. a) The local community structure at a given time step in the vicinity of a randomly selected node in case of the co-authorship network. b) The same picture in the phone-call network. c) The black symbols correspond to the average size of the largest subset of members with the same zip-code, $<n_{real}>$, in the phone-call communities divided by the same quantity found in random sets, $<n_{rand}>$, as the function of the community size $s$. Similarly, the white symbols show the average size of the largest subset of community members with an age falling in a three year time window, divided by the same quantity in random sets. The error-bars in both cases correspond to $<n_{real}>/(<n_{rand}>+\sigma_{rand})$ and $n_{real}>/(<n_{rand}>-\sigma_{rand})$, where $\sigma_{rand}$ is the standard deviation in case of the random sets. d) The $<n_{real}>/s$ as a function of $s$, for both the zip-code (black symbols) and the age (white symbols). e) Possible events in the community evolution. f) The identification of evolving communities. The links at $t$ (blue) and the links at $t+1$ (yellow) are merged into a joint graph (green). Any CPM community at $t$ or $t+1$ is part of a CPM community in the joined graph, therefore, these can be used to match the two sets of communities.

**Figure 2. Characteristic features of community evolution**. a) The age $\tau$ of communities with a given size (number of people) $s$, averaged over the set of available communities and the time steps, divided by the average age of all communities $<\tau>$, as the function of $s$. The increasing nature of the plot indicates that larger communities are on average older. b) The auto-correlation function $C(t)$ of communities with different sizes averaged over the communities and $t_0$. The unit of time, $t$, is two weeks, thus, for the co-authorship network, where the data samples were taken monthly, the $C(t)$ values are shown for every other time step. c) The life-span $\tau^*$ averaged over the communities as the function of the stationarity $\zeta$ and the community size *s* for the co-authorship network. (The communities still living at the last available time step in the dataset were excluded from this investigation). The peak in $<\tau^*>$ is close to $\zeta = 1$ for small sizes, whereas it is shifted towards lower $\zeta$ values for large sizes. d) Similar results found in the phone-call network. In panels (c) and (d) the highlighted white line corresponds to the optimal stationarity.

**Figure 3. Evolution of four types of communities in the co-authorship network.** The height of the columns corresponds to the actual community size, and within one column the yellow colour indicates the number of "old" nodes (that have been present in the community at least in the previous time step as well), while newcomers are shown with green. The members abandoning the community in the next time step are shown with orange or purple colour, depending on whether they are old or new. (This latter type of member joins the community for only one time step). From top to bottom, we show a small and stationary community (a), a small and non-stationary community (b), a large and stationary community (c) and, finally, a large and non-stationary community (d). A mainly growing stage (two time steps) in the evolution of the latter community is detailed in panel e).

**Figure 4. Effects of links between communities.** a) The probability $p_l$ for a member to abandon its community in the next step as a function of the ratio of its aggregated link weights to other parts of the network ($w_{out}$) and its total aggregated link weight ($w_{in}+ w_{out}$). The inset shows the average time spent in the community by the nodes, $<\tau_n>$, in function of $w_{out}/(w_{in}+ w_{out})$. b) The probability $p_d$ for a community to disintegrate in the next step in function of the ratio of the aggregated weights of links from the community to other parts of the network ($W_{out}$) and the aggregated weights of all links starting from the community ($W_{in}+ W_{out}$). The inset shows the average life time $<\tau^*>$ of communities as a function of $W_{out}/(W_{in}+ W_{out})$.

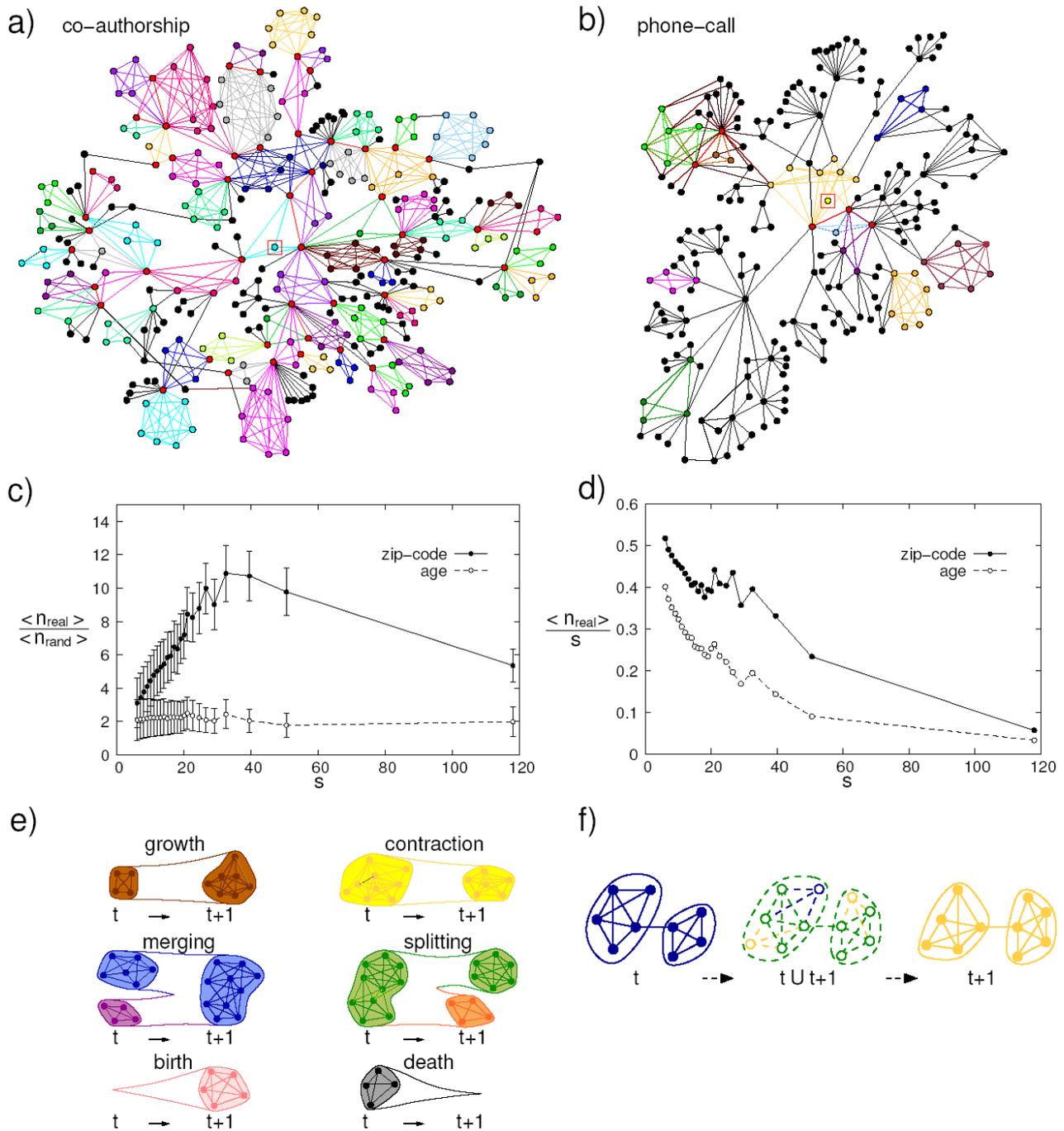

**Figure1.**

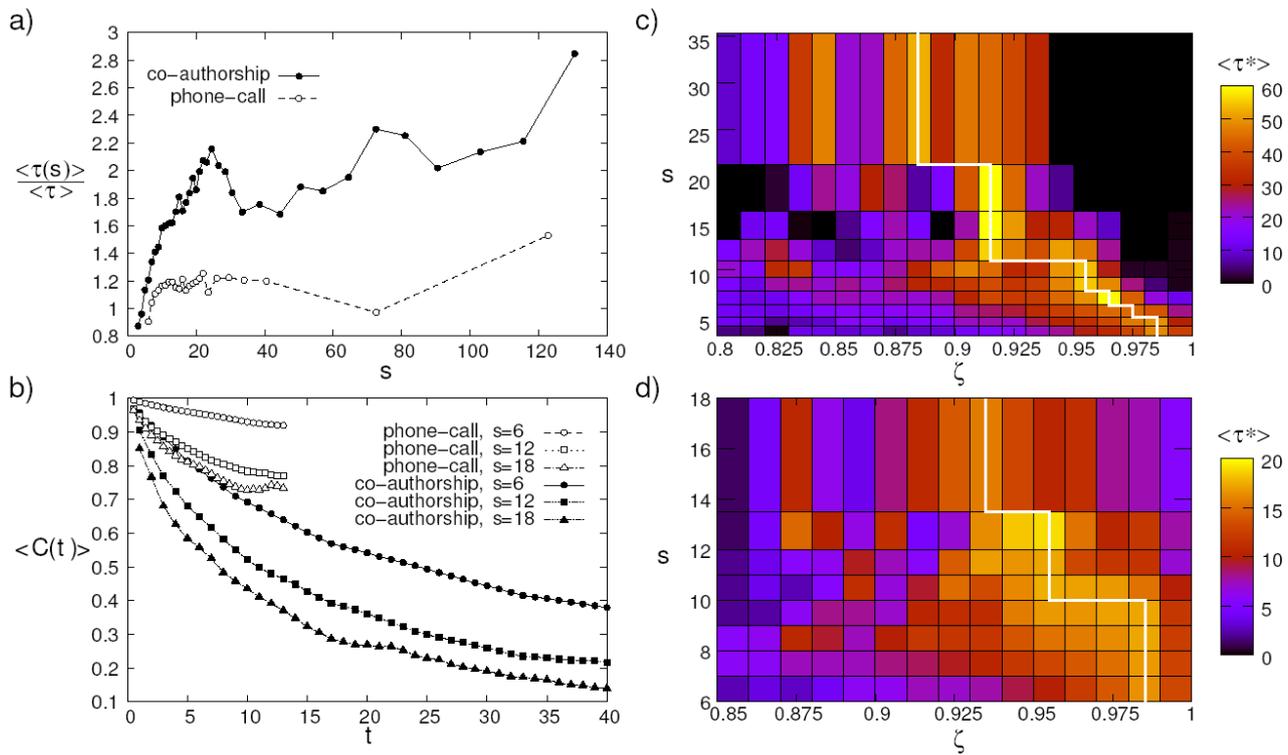

**Figure 2**.

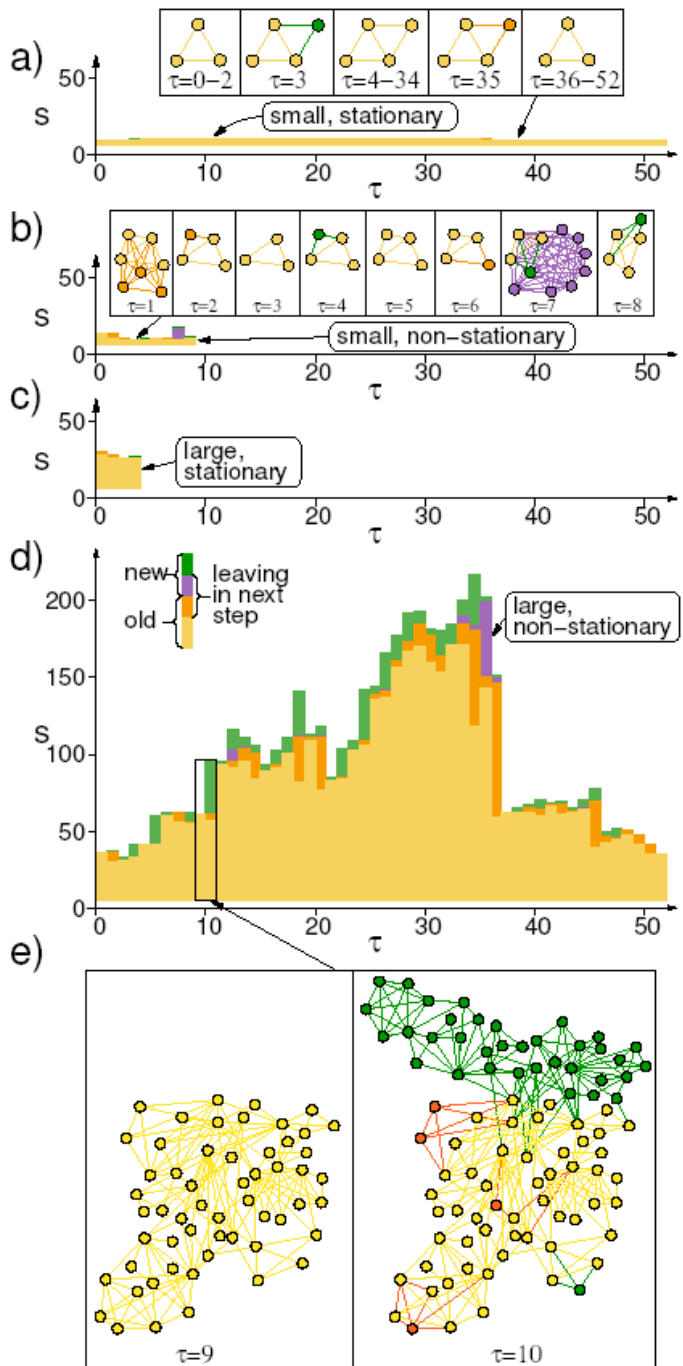

**Figure 3.**

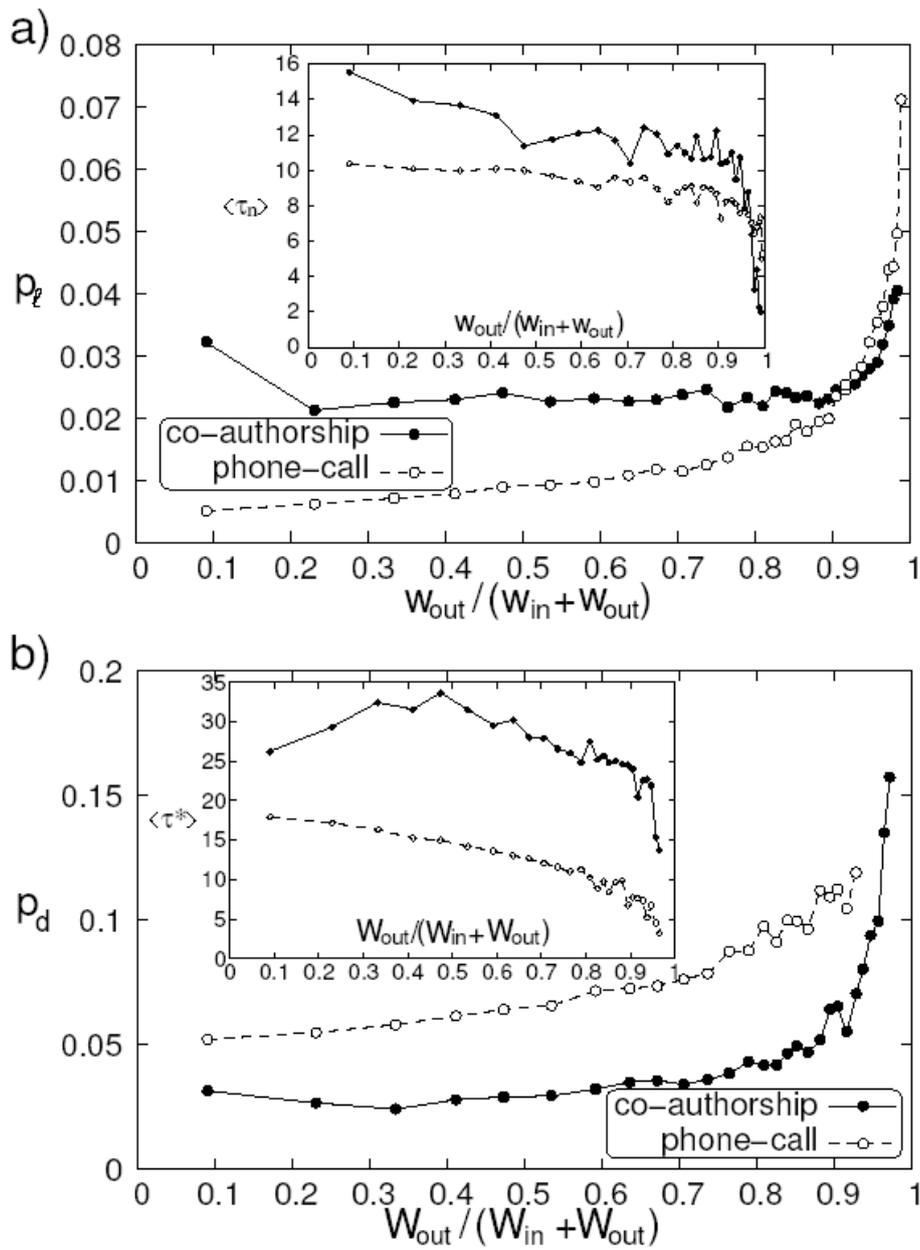

**Figure 4**.